\documentclass[12pt,preprint]{aastex} 
\usepackage{emulateapj5} 
 
\shorttitle{Raman Scattered He~II 6545 in V1016 Cyg}
\shortauthors{Lee et al.}
  
\begin{document} 
   
\title{Raman Scattered He~II $\lambda$ 6545 Line in the Symbiotic Star
V1016~Cygni}

\author{Hee-Won Lee\altaffilmark{1,4}, Young-Jong Sohn\altaffilmark{2,4}, 
Young Woon Kang\altaffilmark{1} \& Ho Il Kim\altaffilmark{3}}
\altaffiltext{1}{Department of Astronomy and Space Sciences, Sejong University, 
       Seoul, 143-747, Korea}
\altaffiltext{2}{Center for Space Astrophysics, Yonsei University, 
       Seoul, 120-749, Korea}
\altaffiltext{3}{Korea Astronomy Observatory, Daejon, 
       305-348, Korea}
\altaffiltext{4}{Visiting astronomer, Cananda-France-Hawaii Telescope 
Corporation,
operated by the National Council of Canada, the Centre National de la
Recherche Scientifique de France and the University of Hawaii}

\begin{abstract}
We present a spectrum of the symbiotic star V1016~Cyg observed with
the 3.6 m Canada-France-Hawaii Telescope, in order to illustrate
a method to measure the covering factor of the neutral
scattering region around the giant component with respect to the
hot emission region around the white dwarf component.
In the spectrum, we find broad wings around H$\alpha$ and a broad 
emission feature around 6545${\rm\ \AA}$ that is blended with the
[N~II]$ \lambda$ 6548 line.  
These two features are proposed to be formed by Raman scattering by atomic
hydrogen, where the incident radiation is proposed to be UV continuum 
radiation around Ly$\beta$ in the former case and 
He~II~$\lambda$~1025 emission line arising from  $n=6\rightarrow n=2$ 
transitions for the latter feature.
We remove the H$\alpha$ wings by a template Raman scattering 
wing profile and subtract the [N~II]~$\lambda$~6548
line using the 3 times stronger [N~II]~$\lambda$~6583 feature in order to 
isolate the He~II Raman scattered 6545 \AA\ line.
We obtain the flux ratio $F_{6545}/F_{6560}=0.24$ of the He~II~$\lambda$~6560 
emission line and the 6545 \AA\ feature for V1016~Cyg. 
Under the assumption that the He~II emission from this object is isotropic, 
this ratio is converted to the ratio $\Phi_{6545}/\Phi_{1025}=0.17$ 
of the number of the incident photons and that of
the scattered photons. This implies that the scattering region with
H~I column density $N_{HI}\ge 10^{20}{\rm\ cm^{-2}}$
covers 17 per cent of the emission region.  
By combining the presumed binary period $\sim 100$ yrs of this system
we infer that a significant fraction of the slow stellar wind from the
Mira component is ionized and that the scattering region around 
the Mira extends
a few tens of AU, which is closely associated with the mass loss process of
the Mira component.
It is argued that the Raman scattered He~II~6545 line is an important
and useful tool to investigate the mass loss process occurring in
the late stage of stellar evolution. 

\end{abstract} 

\section{Introduction}

V1016 Cygni is known to be a symbiotic star consisting of a Mira variable
and a white dwarf. Mira variables usually exhibit a heavy mass loss
with a typical value ${\dot M}= 10^{-7}{\rm\ M_\odot\ yr^{-1}}$ in the
form of a slow stellar wind with a wind speed $\sim 10{\rm\ km\ s^{-1}}$.
V1016 Cyg showed an outburst in 1965, which appears to be a slow nova-like
eruption, indicative of a mass transfer process between the Mira and
the white dwarf (Fitzgerald et al. 1966). The outburst is proposed to be
thermonuclear runaway on the surface of the white dwarf component.

In symbiotic stars, strong UV radiation, which is  produced 
around the mass accreting 
white dwarf, ionizes the surrounding nebula giving rise to prominent 
emission lines (e.g. Kenyon 1986). 
Furthermore, a significant part of a slow stellar wind
from the Mira component will be ionized, whereas  a
predominantly neutral region may reside in the vicinity of the giant star.
Therefore, useful information about binary interaction may be obtained from
an estimate of the extent of the neutral region around the giant.

About a half of the symbiotic stars including V1016 Cyg are known to exhibit
broad emission features around 6825 \AA\ and 7088 \AA, which Schmid (1989)
identified
as the Raman scattered O~VI $\lambda\lambda$ 1032, 1038 by atomic
hydrogen. The Raman scattering process starts with a scattering hydrogen
atom initially in the $1s$ state and an incident UV photon redward of
Ly$\beta$ and ends up
with the hydrogen atom in the $2s$ state and an optical photon redward of 
H$\alpha$. The operation of this Raman scattering process requires the
co-existence of a hot ionized O~VI emission nebula and 
a cold neutral scattering region, which is plausibly satisfied in the
case of symbiotic stars.
Simultaneous detections of the features in the far UV and optical
regions support strongly the Raman scattering nature of these emission
features (Birriel, Espey \& Schulte-Ladbeck 1998, 2000).

High resolution spectroscopy and spectropolarimetry show that
these Raman scattered O~VI features exhibit multiple peak structures
and strong polarization often accompanied by polarization flip in the red part  
(Schmid \& Schild 1994, Harries \& Howarth 1996, 2000). 
Whereas Raman scattering of O~VI 1032, 1038 operates in the
presence of neutral H~I scattering regions with H~I column density
$N_{HI}\sim 10^{22}{\rm\ cm^{-2}}$,  UV incident radiation much closer
to Lyman series is scattered in a region with much less
$N_{HI}$. Examples of these processes include the formation of broad
H$\alpha$ wings and He~II Raman scattered lines.

Nussbaumer, Schmid \& Vogel  (1989) advanced an idea that 
H$\alpha$ wings can be made through Raman scattering of Ly$\beta$, 
which was applied in a quantitative way by Lee \& Hyung(2000) in their
analysis of the H$\alpha$ wings of the planetary nebula 
IC~4997. Many symbiotic stars also exhibit similar broad H$\alpha$
wings that are fitted very well with the template
wing profiles obtained from Raman scattering of Ly$\beta$ 
(Lee 2000, see also Ivison, Bode, \& Meaburn 1994, 
van Winckel, Duerbeck, \& Schwarz 1993). 

With their recent extensive spectroscopy of young
planetary nebulae and post AGB stars, Arrieta \& Torres-Peimbert (2003) 
found that H$\alpha$ lines exhibit broad wings in these objects. 
They showed that the wings are well fitted by a
simple formula proportional to $\Delta\lambda^{-2}$, 
giving  support for the Raman
scattering origin of H$\alpha$ wings. 
Broad wings around H$\alpha$ are also reported in many post AGB objects
(e.g. Balick 1989, Selvelli \& Bonafacio 2000, van de Steene,
Wood \& van Hoof 2000, Arrieta \& Torres-Peimbert 2000).

A He~II ion being a single electron atom, He~II emission lines arising
from the transitions between energy levels with even principal 
quantum numbers have very close wavelengths to those for
the H~I resonance transitions. 
Van Groningen (1993) found a broad feature around 4851 \AA\ 
in the spectrum of the symbiotic nova RR Telescopii, 
which he proposed to be formed through Raman scattering of He~II
emission lines by atomic hydrogen. Raman scattered He~II may be observed 
in broader classes of objects than symbiotic stars.  
In particular, in the planetary nebula NGC~7027, 
P\'equignot et al. (1997) reported the existence of
He~II Raman-scattered feature around 4851 \AA\  blueward of H$\beta$.  
He~II Raman-scattered 6545 line was also found in the symbiotic stars
RR Tel and He2-106 (Lee, Kang \& Byun 2001). 

In this paper, we present a spectrum of the symbiotic star V1016 Cyg 
observed with
the 3.6 m Canada-France-Hawaii Telescope,
from which we isolate the He~II Raman scattered 6545 \AA\ feature.
We calculate the ratio of the number fluxes of
incident and scattered radiation in a straightforward way, 
which will provide a direct measure of
the covering factor of the neutral scattering region.

\section{Data and Analysis}
We observed the symbiotic star V1016 Cyg on the night May 20, 2002
with the coud\'e spectrograph `Gecko'  
at the 3.6 m Canada-France-Hawaii Telescope at Mauna Kea. 
A 2k$\times$ 4.5k EEV1 CCD chip was used for the detector,
and the order sorting was made with an interference filter.
The spectral resolution was $R=120,000$.
Using IRAF packages, standard procedures have been followed 
in order to reduce the data. 

In Fig.~1, we show the spectrum of V1016 Cyg.
We show fluxes in the vertical axis with arbitrary unit, because
in this work the only necessary quantities are relative ratios.
The vertical arrows in the figure mark 
He~II~$\lambda$~6527, the Raman scattered He~II~6545 \AA, and
He~II~$\lambda$~6560. We blow up the data by a factor of 50 in order to see
clearly the weak He~II~$\lambda$~6527 and the 6545 feature represented
by the solid thick line. 

The emission line He~II~$\lambda$~6527 arises from
a transition between $n=14$ and $n=5$ states of He~II.
Raman scattered He~II 6545 \AA\ feature is formed when a
He~II~$\lambda$~1025 line photon ($n=6\rightarrow n=2$) is incident upon a 
hydrogen atom in the ground
state followed by a de-excitation of the scattering hydrogen atom into
$2s$ state re-emitting an optical photon with $\lambda=6545$ \AA.  
Because of the energy conservation, the wavelength $\lambda_o$
of the Raman scattered radiation is associated with that of the
incident UV radiation by
\begin{equation}
\lambda_o=(\lambda_i^{-1}-\lambda_{Ly\alpha}^{-1})^{-1},
\end{equation}
where $\lambda_{Ly\alpha}$ is the wavelength of hydrogen Ly$\alpha$.
The variation $\Delta\lambda_i$ in the incident radiation results in
corresponding variation of $\Delta\lambda_o$ given by
\begin{equation}
{\Delta\lambda_o \over\lambda_o}
={\lambda_o\over\lambda_i}{\Delta\lambda_i\over\lambda_i}
\end{equation}
which leads to the broadening of the Raman scattered radiation 
by a factor of 6.4.

He~II emission lines from $2n\rightarrow 2$ transitions
have wavelengths that differ from those of H~I Lyman series lines
by 1/2000 due to the small difference in reduced masses. This results
in Raman-scattered radiation with wavelength differences 
from Balmer lines
\begin{equation}
\Delta\lambda_o  
\simeq  5.9\left[{n^2(n^2-1)\over (n^2-4)^2}\right] {\rm\ \AA} ,
\end{equation}
(Lee et al. 2001).

The scattering cross section for this process can be computed using the
Kramers-Heisenberg formula, which yields $\sigma\sim 10^{-20}{\rm\ cm^{-2}}$.
(e.g. Lee et al. 2001, Nussbaumer et al. 1989). 
This cross section is two orders of magnitude larger than the
corresponding values for the Raman scattering of O~VI 1032, 1038.
Fig.~2 shows the energy level diagrams of hydrogen and He~II, where
emission lines of He~II and the Raman
scattering processes are illustrated.

In this spectrum, there appear very broad wings around H$\alpha$, which
are excellently fitted in the far wing parts 
by $\Delta\lambda^{-2}=(\lambda-\lambda_{H\alpha})^{-2}$ 
profile represented by a dotted line in Fig.~1,
$\lambda_{H\alpha}$ being the difference of the
wavelength from that of H$\alpha$. This profile is proposed to be
originated from Raman scattering of flat incident radiation around
Ly$\beta$ by Lee (2000) and Lee \& Hyung (2000). We prepare the template
profiles of Raman scattering wings that are expected to be formed 
in neutral scattering regions with H~I column densities
$N_{HI}=10^{19}-10^{21}{\rm\ cm^{-2}}$
irradiated by a flat continuum radiation around Ly$\beta$, as was
done by Lee \& Hyung (2000). These template wings
are shown in Fig.~3. The exact choice of the H~I column density does not
alter the main result of this work, because it affects the shape near
the H$\alpha$ center and we are only interested
in the relative strength ratio of He~II emission and Raman-scattered
lines that lie in the far blue wing part of H$\alpha$. 
We select the template wing profile for $N_{HI}=10^{21}{\rm\ cm^{-2}}$,
which is also capable of producing  a Raman scattered He~II 6545 feature.
 
After subtraction of the H$\alpha$ wings, we perform Gaussian fittings
with a functional form $f(\lambda)=f_0 
\exp[-(\lambda-\lambda_0)^2/\Delta\lambda^2]$
to He~II emission lines at 6527 \AA\ and 6560 \AA, and the H$\alpha$, using
a least square method. Satisfactory fits are obtained by a single Gaussian
for He~II emission lines at 6527 \AA\ and 6560 \AA\ and H$\alpha$.
The line widths are 
$\Delta\lambda = 0.96 \ {\rm\AA}, 0.98\ {\rm \AA}$ for He~II 6527 and 6560
respectively, and for H$\alpha$ we obtain $\Delta\lambda=1.0 {\rm \AA}$.  
The full widths at half maximum of these 
emission lines are $73{\rm\ km\ s^{-1}}$, $74{\rm\ km\ s^{-1}}$ and
$78{\rm\ km\ s^{-1}}$, respectively.  The peak values are 
$f_0=5.2\times 10^4$ for He~II~$\lambda$~6560 and $f_0=1.8\times 10^3$ 
for He~II~$\lambda$~6527.  Fig.~4 shows our results of Gaussian
fittings for H~I and He~II emission lines.  

In order to separate the Raman scattered He~II 6545 feature from the
spectrum,  we make use of the fact that
the [N~II]~$\lambda$~6548 profile should be identical with that
of the [N~II]~$\lambda$~6584 feature, which is 3 times stronger
due to the atomic structure (e.g. Osterbrock 1989, Storey \& Zeippen 2000).
We subtracted the flux of the [N~II]~$\lambda$~6584
multiplied by 1/3 (represented by red lines) from the flux around the 
6545 and [N~II]~$\lambda$~6548 features (represented by the black lines)
to isolate the He~II Raman-scattered feature.

We show the result in Fig.~5, where the Raman scattered He~II~6545 feature 
is seen clearly
in solid thick line. The dotted line is a Gaussian fit to the feature,
where the width is $\Delta\lambda = 6.2{\rm\ \AA}$ and the peak value is
$f_0=1.0\times 10^3$. This width is simply chosen
from the mean value of the widths of the He~II emission lines 
multiplied by a factor 6.4, which is 
the enhancement factor of the scattered line width 
attributed to the scattering incoherency. 

\section{Calculation}

\subsection{Strength of Raman Scattered He~II~6545 Line}

The He~II~$\lambda$~1025 flux of V1016 Cyg can only be indirectly
inferred from the fluxes of He~II emission lines at 6527~\AA\
and 6560~\AA. From the recombination theory, relative strengths
of He~II emission lines can be calculated as a function of temperatures and
electron densities (e.g. Storey \& Hummer 1995, Brocklehurst 1971).  
According to Gurzadyan (1997) the calculated fluxes of 
He~II~$\lambda$~6560 and He~II~$\lambda$~1025 lines are given by
$F_{6560}=0.135 F_{4686},\quad F_{1025}=0.618 F_{4686}$, where $F_{4686}$ 
is the flux of He~II 4686. 
In this work, it is assumed that the temperature is 
$T=20000 {\rm\ K}$ and the electron density $n_e=10^4{\rm\ cm^{-3}}$ for
illustrative purpose. If we consider the case B recombination, 
the ratio of the number fluxes of the He~II~$\lambda$~6560 and 
He~II~$\lambda$~1025 line photons 
generated from recombination is 
\begin{equation}
\Phi_{1025}/\Phi_{6560}=(0.618/0.135) (1025/6560) = 0.715.
\end{equation}
This implies that the number of He~II~$\lambda$~1025 line photons is almost
70 per cent that of He~II~$\lambda$~6560 line photons, 
which does not vary sensitively
with the physical conditions assumed above. Because the Raman He~II 6545
is located close to He~II~$\lambda$~6560, their flux ratio directly indicates
the scattering efficiency.

Under the assumption that the He~II emission from this object is isotropic, 
this ratio is converted to the ratio of the number of the incident 
photons and that of the scattered photons 
\begin{equation}
\Phi_{6545}/\Phi_{1025}={1.1\times 10^3\over 5.2\times 10^4}\ {6545\over
1025}\ {\Phi_{1025}\over\Phi_{6560}}=0.17. 
\end{equation}
Birriel et al. (2000) 
used the Hopkins Ultraviolet Telescope (HUT) and various ground based 
telescopes to measure the strengths of the resonance 
doublet O~VI 1032, 1038 and their Raman scattered features at 6825 \AA and 
7088 \AA, in order to obtain the Raman scattering efficiencies in a number of
symbiotic stars. They conclude that the scattering 
efficiency for V1016 Cyg is about 50 \%, which is larger than our
own estimate.

\subsection{Neutral Scattering Region and Mass Loss Process of the Mira} 

\subsubsection{Covering Factor of the Neutral Scattering Region}

In this subsection we consider the neutral scattering
region in V1016 Cyg which is responsible for the formation of the 6545
feature.
The giant component of V1016 Cyg is believed to be a Mira variable,
and the white dwarf component showed outburst in 1965. 
A Mira variable usually exhibits a mass loss at a rate ${\dot M}\sim
10^{-4}-10^{-7} {\rm\ M_\odot\ yr^{-1}}$ in the form of a slow stellar wind
with a typical wind velocity $v\sim 10{\rm\ km\ s^{-1}}$ (e.g. Vassiliadis
\& Wood 1993, Schild 1989).  
The neutral scattering region is plausibly formed around the
giant component and closely related with the mass loss process of the
giant. 

The mass loss rate of a Mira variable may not be a continuous process 
and can be episodic as it evolves
in the asymptotic giant branch, which was proposed by Zijlstra et al. (1992).
However, for simplicity, we assume 
that the mass loss rate is constant and the mass loss process is
spherically symmetric. In a binary system like V1016 Cyg, this
assumption may be quite unrealistic because the material distribution will
be severely affected by the presence of the white dwarf companion and
the orbital motion. 
Assuming a wind consisting of only hydrogen we may write the mass loss
rate
\begin{equation}
{\dot M}=4\pi r^2 m_p n(r) v(r).
\end{equation}

The slow stellar wind velocity is almost constant in a quite extended
region, and therefore the density is expected to decrease as $r^{-2}$.
If we assume that this spherically symmetric region consists of
complete neutral hydrogen, any sight line originating from the emission region
around the white dwarf will have the Raman
scattering optical depth exceeding unity 
for a mass loss rate of ${\dot M}=10^{-7}{\rm\ M_\odot\ yr^{-1}}$. 
Therefore, we may expect that the wind material should be ionized 
significantly. The 
ionization front is generally sharply defined on the order of the mean free
path of Lyman limit photons, which is quite small compared with the
characteristic scale of the scattering region.

Schmid (1995) considered the distribution of neutral hydrogen around the
giant component when he investigate Rayleigh and Raman scattering processes.
In his computation, he introduced a dimensionless parameter $X_{H^+}$ 
defined by
\begin{equation}
X_{H^+}\equiv {4\pi\mu^2 m_p^2\over a_B(H,T_e)a_H^2}
DL_H\left({v_\infty\over \dot M}\right)^2.
\end{equation}
Here, $a_B(H,T_e)$ is the recombination coefficient of hydrogen in a
region with electron temperature $T_e$, $D$ is the binary separation,
$L_H$ is the number
of ionizing photons from the hot white dwarf, $\mu$ is the mean
molecular weight and $a_H$ is the relatvie hydrogen particle density.

For given $\theta$ the distance $s=uL$ to the ionization front 
from the He~II emission region is computed by considering the
integral given by
\begin{eqnarray}
f(\theta, s) &=& \int_0^u {x^2 dx \over (x^2-2x\cos\theta+1)^2}
\nonumber \\
&=&{1\over\sin^3\theta}\left[{1\over2}\left(\tan^{-1}{u-\cos\theta\over
\sin\theta}-\theta+{\pi\over2} \right) \right. \nonumber \\
&+& \left. {1\over 4}\sin 2\left(\tan^{-1}{u-\cos\theta\over\sin\theta}
-\theta\right)\right].
\end{eqnarray}
The ionization boundary can be located
when $X_{H^+}$ exceeds $f(\theta)$ defined by
\begin{eqnarray}
f(\theta)&=& \lim_{s\rightarrow\infty}f(\theta,s)
\nonumber \\
&=&{1\over\sin^3\theta}\left({\pi-\theta\over2}+{\sin2\theta\over4}
\right).
\end{eqnarray}
Fig.~6 shows the graph of $f(\theta)$. As is pointed by Schmid (1995), when
$X_{H^+}>{\pi/4}=f(\theta=\pi/2)$, most of the giant wind is
ionized leaving a cone-like neutral region around the giant.

It is expected that the sight line from the He~II emission region
will be characterized by a large H~I column density $N_{HI}> 10^{21}
{\rm\ cm^{-2}}$, at which the Raman scattering optical depth becomes 
unity. Therefore, we may safely regard that He~II~$\lambda$~1025 line photons
will be Raman-scattered with almost same efficiency.
Lee \& Lee (1997) performed detailed Monte Carlo calculations
to show that the efficiency is about 0.6, which means that 
about 60 per cent of He~II~$\lambda$~1025 photons that enter a neutral region 
with the Raman scattering optical depth exceeding unity will come out as 
Raman-scattered optical photons. 
The remaining 40 per cent of UV photons are reflected near the surface by
Rayleigh scattering. Since 17 per cent of He~II~$\lambda$~1025
photons are converted into Raman scattered He~II 6545 photons,
it follows that the covering factor $C_S$ of the scattering region
is $C_S \sim 0.17/0.6 = 0.28$.
Noting that
\begin{equation}
C_S ={\Delta\Omega \over 4\pi} ={1-\cos\theta_S \over 2}
\end{equation}
we obtain $\theta_S =64^\circ$, where $\Delta\Omega$ is the solid angle
of the scattering region subtended with respect to the He~II emission
region.  
Substituting $\theta_S=64^\circ$ into Eq.~(9), we obtain
\begin{equation}
X_{H^+}=f(\theta_S)\simeq 1.9.
\end{equation}
This estimate is similar to the value $X_{H^+}\sim 2.3-3.7$ proposed
by M\"urset et al. (1991), but quite large compared with the value
$X_{H^+}\sim 0.4$ obtained by Birriel et al. (2000).

\subsubsection{Mass Loss Rate of the Mira}

Despite intensive research over several decades, the orbital period
of V1016 Cyg is only poorly known. 
The distance to V1016 Cyg is also highly uncertain with estimates ranging
from 1 kpc to 10 kpc (Watson et al. 2000 and references therein).
From a recent analysis using optical images obtained
with the Hubble Space Telescope and radio images from 
the Vary Large Array (VLA),
Brocksopp et al. (2002) suggested a projected binary separation of
$84\pm 2{\rm\ AU}$ with the assumption of a distance 2 kpc. Using this
separation, they proposed that the binary period will be longer than
500 years. Previous works by other researchers provide smaller binary
separations and shorter binary periods. For example, Schmid \& Schild
(2002) performed spectropolarimetry to investigate the variation of
the polarization direction which is expected to vary due to the
orbital motion. Based on their polarimetric method, they proposed that
the binary period may exceed 100 years.

Here, we assume that the distance to V1016 Cyg is 2 kpc as Brocksopp et al.
(2002) did. With the H~I distribution given in Eq.~(6), the H~I
column density for a sight line into the neutral region is 
nearly inversely proportional to the impact parameter $p$,
\begin{equation}
N_{HI}\simeq \left({\pi\over 2} \right) n_*R_* \left({R_*\over p}\right) 
\end{equation}
where $R_*, n_*$ is the radius of the Mira and the hydrogen density near
the  surface of the Mira. 
Assuming the mass loss given in Eq.~(6) we have
\begin{eqnarray}
n_* R_*\left({R_*\over p}\right) 
&=&2.5\times 10^{21}{\rm\ cm^{-2}}\left({10{\rm\ AU}\over p} \right)
\nonumber \\
& &
\left({{\dot M_\odot} \over 10^{-7}{\dot{\rm M_\odot}\ {\rm yr^{-1}}}}\right)
\left({10{\rm\ km\ s^{-1}}\over v_*} \right)
\end{eqnarray}

The impact parameter $p$ for the sight line that grazes the ionization
boundary with $\theta=\theta_S$, we have $p=D\sin\theta_S=0.9 D$, where
$D$ is the binary separation of V1016 Cyg.
The Raman scattering optical depth for this sight line 
exceeds unity, so that $N_{HI}\ge 0.9\times 10^{21}{\rm\ cm^{-2}}$. Therefore,
we have
\begin{equation}
{\dot M} \ge 10^{-7}
\left({D \over 25{\rm\ AU} }\right)
\left({v_*\over 10{\rm\ km\ s^{-1}}} \right){\rm\ M_\odot\ yr^{-1}}.
\end{equation}

If we adopt the binary separation $D=80{\rm\ AU}$ 
obtained by Brocksopp et al. (2002), and assume the stellar wind
velocity about $v_*=10{\rm\ km\ s^{-1}}$ near the Mira surface,
we obtain ${\dot M}\ge 3\times 10^{-7}{\rm\ M_\odot\ yr^{-1}}$.
However, this estimate is highly dependent on the choice of the
distance to V1016 Cyg, which is currently quite uncertain. 

\section{Discussion}

In this paper, we introduced a method
to obtain a covering factor of
the neutral scattering region with respect to the emission region
formed around the white dwarf component in a symbiotic star.
Since the spacing between the Raman scattered 6545 feature 
and He~II~$\lambda$~6560 emission line  is quite small, a single 
exposure is needed to obtain both features and the effects including
interstellar extinction can be avoided as far as we are concerned
with the flux ratio of both features.  Furthermore, the Raman scattered
He~II 6545 feature can be isolated with relative ease from 
[N~II]~$\lambda$~6548 feature using always 3 times stronger 
[N~II]~$\lambda$~6584 line, which makes the current method quite robust.
With this ratio, we may obtain important information regarding
the ionization radiation and the mass loss process in symbiotic stars.

Both H$\alpha$ broad wings  and the 6545 feature have Raman scattering origin,
where continuum around Ly$\beta$ is responsible for the former feature
and He~II~$\lambda$~1025 emission line is for the latter. Therefore, the 
strength ratio of the 6545 feature relative to that of H$\alpha$ wings 
is equal to the flux ratio of the He~II~$\lambda$~1025 line to 
that of the continuum around
1025 \AA. Since the H$\alpha$ wings are approximated by $1.4\times 10^5 
(\lambda-\lambda_{H\alpha})^{-2}$, and the 6545 feature is fitted with
the Gaussian $1.12\times 10^3 e^{-[(\lambda-6545\AA)/(5\AA)]^2}$,
the equivalent width of the 6545 feature
\begin{equation}
EW_{6545}={1.12\times 10^3\over 473.2} 
\left(\int e^{-(\lambda/5{\rm\ \AA})^2} d\lambda \right) {\rm\ \AA}
 =21{\rm\ \AA}
\end{equation}
Noting that the wavelength scale is enlarged by a factor of 6.4 in
the optical region due to the incoherence of Raman scattering,
we obtain the equivalent width of He~II~$\lambda$~1025 is $EW_{He II 1025}
= 3.3 {\rm\ \AA}$. A direct measurement of this quantity is severely
hindered by the interstellar medium. However, we may infer it by measuring
the equivalent width of nearby He~II 1080 line using FUSE.

Espey et al. (1995) used the data obtained with the HUT to 
estimate that 25-50 per cent of O~VI 1032, 1038 doublet
photons interact in the neutral scattering region in the symbiotic star
RR Tel. Because the distance and the binary orbital period of RR Tel are 
also very uncertain, a direct comparison with the results for V1016 Cyg
is not possible. However, it appears that a very similar scattering geometry
is shared by these two symbiotic novae. Even though He~II~$\lambda$~1025 has
two orders of magnitude larger scattering cross section than O~VI 1032, 1038
lines, it may be that they share almost same neutral scattering region
due to strong H-ionizing UV radiation from the white dwarf component.

Schmid et al. (1999) proposed that the strength of Raman scattered O~VI 6827, 
7088 may change with the binary orbital phase because the scattering
process is a dipole process where forward and backward scattering is
more favored than orthogonal scattering. The same argument also applies 
to the Raman scattered He~II 6545 lines, where the flux of He~II 6545
feature is expected to be stronger in the conjunction phases than
in the quadrature phases. In order to test this expectation, it requires
a very long term monitoring. More information may be obtained from
spectropolarimetry as has been done for Raman scattered O~VI lines 
(e.g. Harries \& Howarth 1996, Schmid \& Schild 1996). Binary orbit parameters
have been extracted in some symbiotic stars by noting the variation
of the position angle. 

There may be additional scattering regions that are not considered in
this work. For example, spectropolarimetry around Raman scattered
O~VI 6827 in many symbiotic stars shows that the polarization direction
in the reddest part with $v\ge +80{\rm\ km\ s^{-1}}$ is 
flipped by an amount of almost $90^\circ$ 
with respect to the strongly polarized blue and main part (e.g.
Harries \& Howarth 1996). Schmid (1996) proposed that this flip
is attributed to the receding part of the slow stellar wind around the
giant component. However, the typical wind speed is $\sim 10 {\rm
\ km\ s^{-1}}$, which is much smaller than the observed values.

Lee \& Park (1998) proposed that the profile of O~VI 6827 is
mainly attributed to the kinematics of the emission region formed around
an accretion disk. The emission region may extend in sub-AU scale with
the Keplerian speed $\sim 100{\rm\ km\ s^{-1}}$ in the orbital plane. 
In this picture, the flipped red part can be explained
by proposing the existence of  additional neutral or clumpy regions that are
moving away in the direction normal to the orbital plane. This raises
an interesting possibility that this additional component is related
with the bipolar morphologies exhibited in many symbiotic stars.
However, this additionally scattered component appears weak in O~VI 6827
features, and maybe this does not contribute much to the He~II 6545 feature.
Future spectropolarimetry using big telescopes may shed more
light about this.

We conclude that the Raman scattered He~II 6545 feature is extremely
useful to investigate the mass loss process of a giant star and
the size of the neutral scattering region,
which will put very tight constraints on the distance and binary
orbital period of symbiotic stars.

\acknowledgements
We thank the staff at CFHT for their support in telescope
operation. The referee, Dr. Birriel, provided helpful comments that improved
the presentation of this paper.  We are also grateful to Jaewoo Lee 
for his careful and useful comments.
YJS acknowledges support provided by the Creative Research Initiative Program
of the Korean Ministry of Science and Technology.
This work was supported by the Astrophysical Research 
Center for the Structure and Evolution of the Cosmos (ARCSEC) funded
by the Korea Science and Engineering Foundation and the Korean Ministry of 
Science and Technology.

\clearpage

\begin{figure}
\plotone{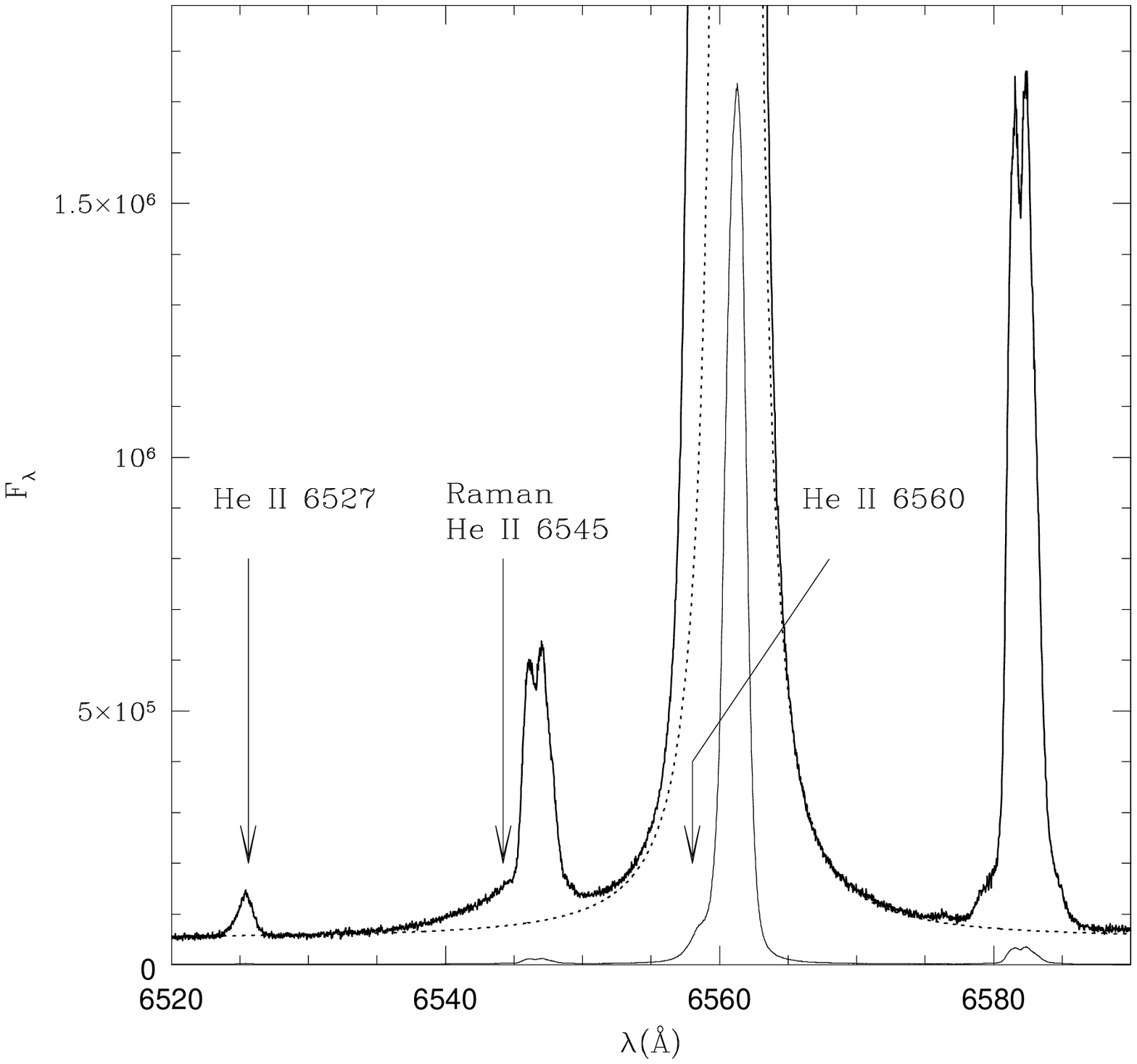} \caption{
The spectrum around H$\alpha$ of the symbiotic Mira V1016 Cyg obtained with
the 3.6 m Canada-France-Hawaii Telescope.  The vertical arrows mark 
He~II~$\lambda$~6527, the Raman scattered He~II 6545 \AA, and
He~II~$\lambda$~6560. The solid thick line is a blow-up of the solid
thin line by a factor of 50, where
the weak He~II $\lambda$~6527 and the 6545 feature are clearly illustrated.
The flux is shown in arbitrary units because physically meaningful
quantities are relative fluxes in the current work.
Note that there appear very broad wings around H$\alpha$, which
are excellently fitted by $\Delta\lambda^{-2}=(\lambda-\lambda_{H\alpha})^{-2}$ 
profile represented by the dotted line,
$\lambda_{H\alpha}$ being the difference of the
wavelength from that of H$\alpha$. }
\end{figure}
\begin{figure}
\plotone{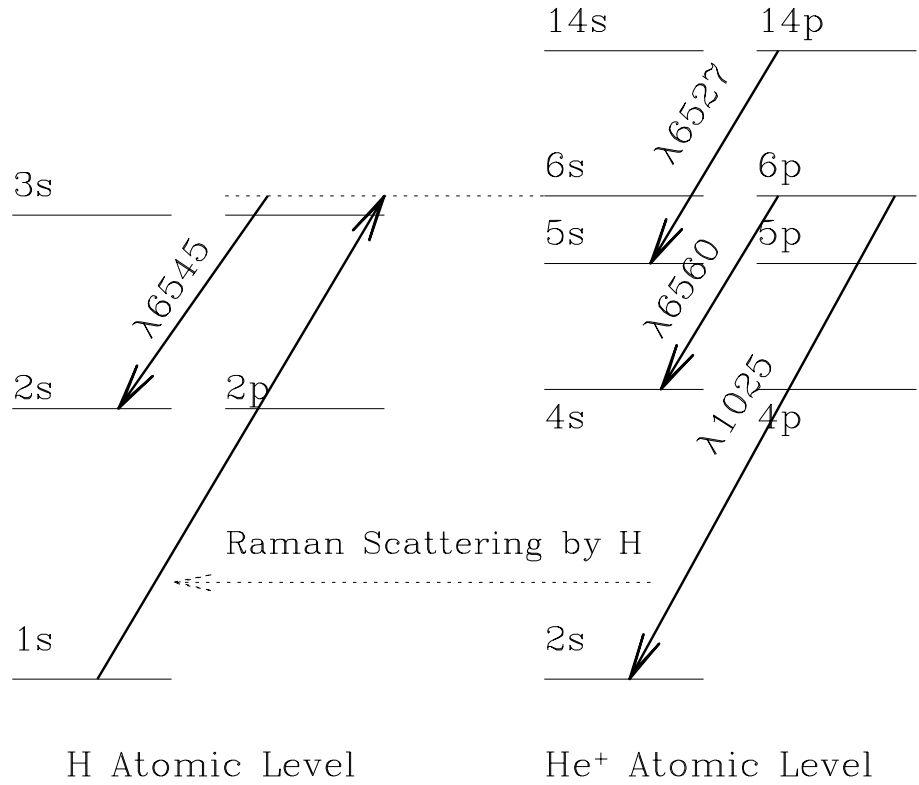} \caption{
Atomic level structures of hydrogen and He~II. He~II emission lines at 
6527 \AA\ and 6560 \AA\ arise from $n=14\rightarrow n=5$ transitions
and $n=6\rightarrow n=4$ transitions, respectively.
He~II~$\lambda$~1025 ($n=6\rightarrow n=2$) emission line has a slightly higher
energy than Ly$\beta$ 1025. When a  He~II~$\lambda$~1025 photon is 
incident upon
a hydrogen atom, it may de-excite to $2s$ state by re-emitting an optical
photon blueward of H$\alpha$ which appears at 6545 \AA. } 
\end{figure}

\begin{figure}
\plotone{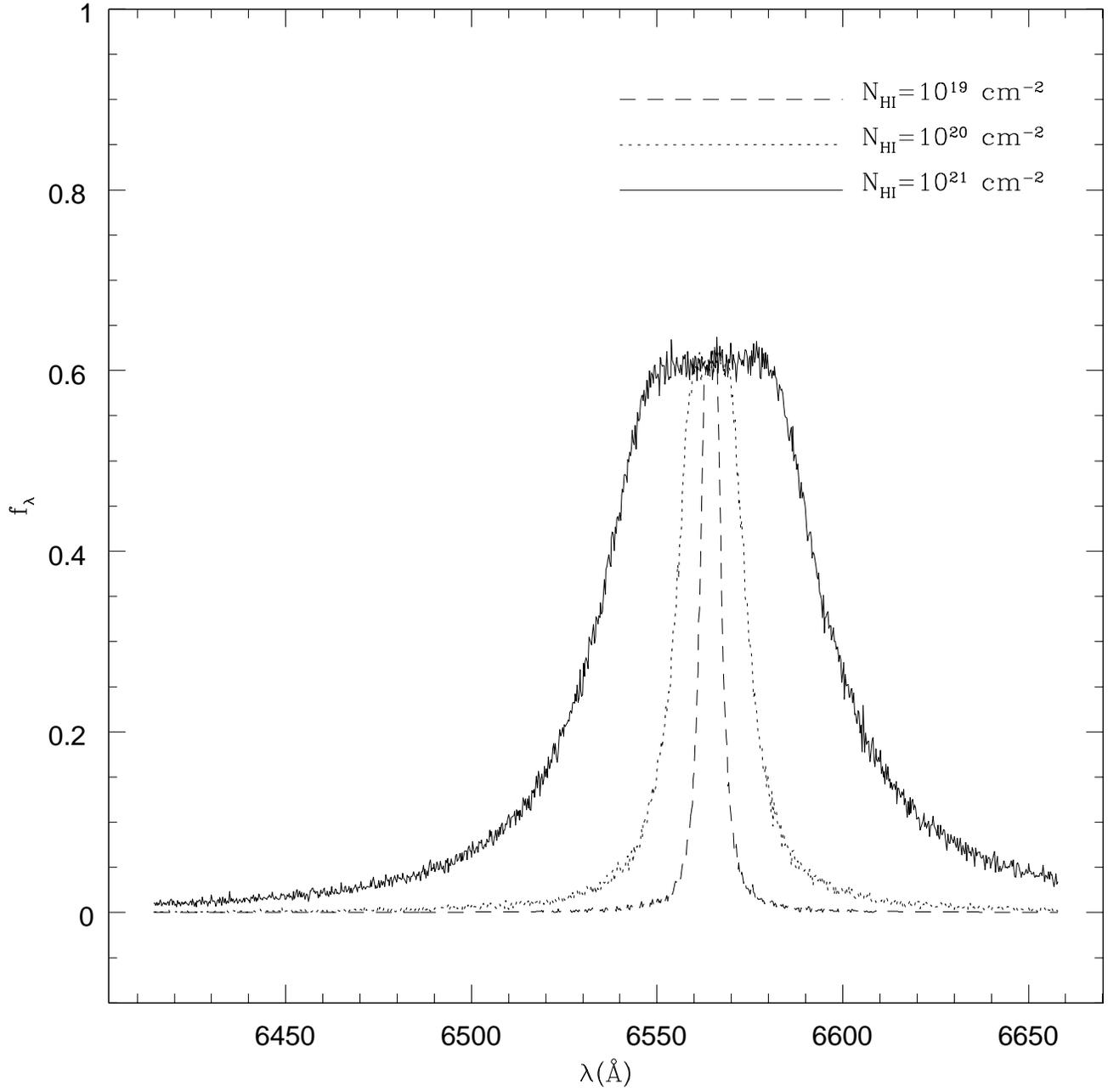} \caption{
Template wing profiles around H$\alpha$ that are formed from
Raman scattering of a flat continuum around Ly$\beta$ in a neutral
scattering region with H~I column density $N_{HI}$.
The wing profiles were prepared using a Monte Carlo technique, of which
the detailed procedure is explained in Lee \& Hyung (2000). }
\end{figure}

\begin{figure}
\plotone{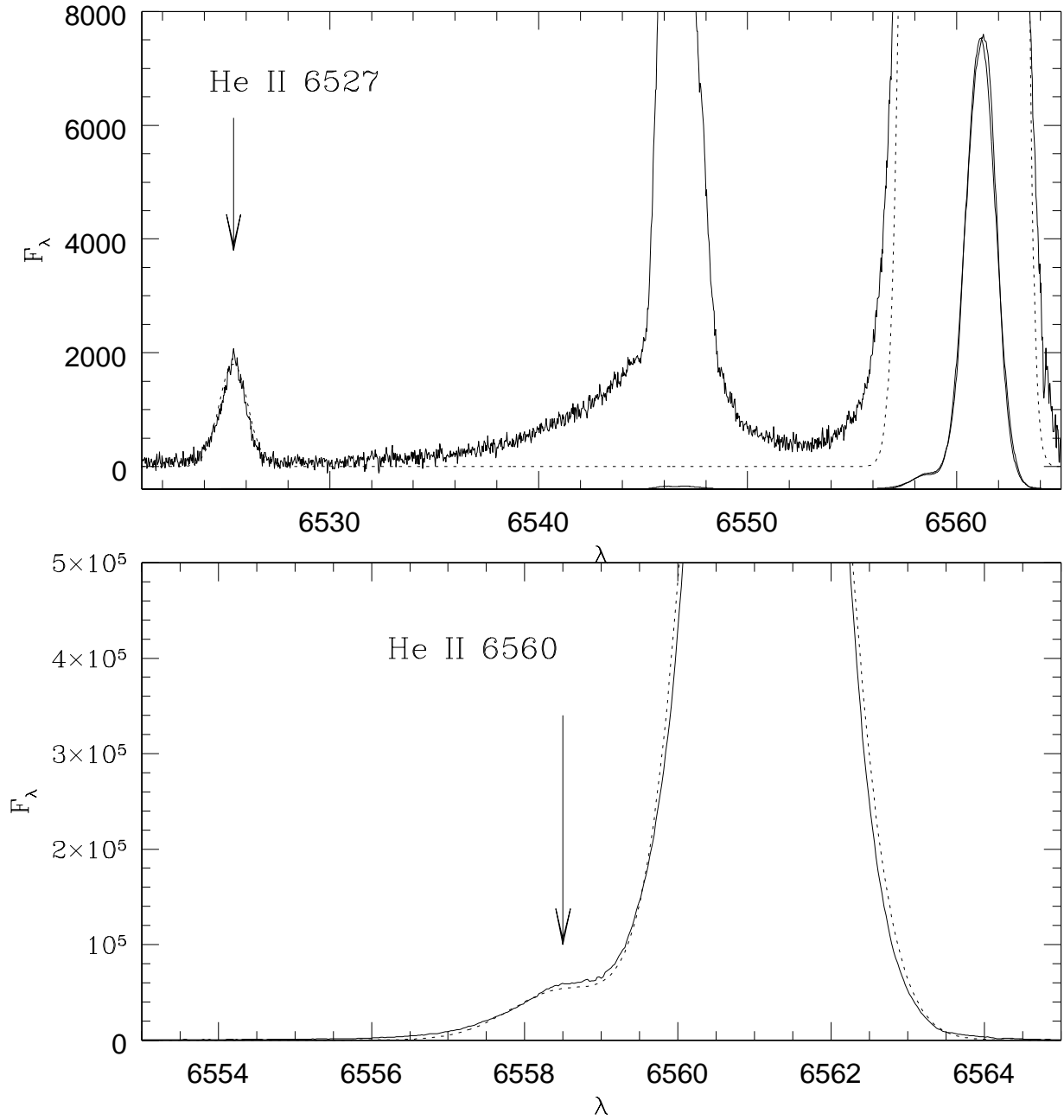} \caption{
Gaussian fits using a functional form $f(\lambda)=f_0 
\exp[-(\lambda-\lambda_0)^2/\Delta\lambda^2]$
to He~II$~\lambda$~6527, He~II~$\lambda$~6560 and 
H$\alpha$ after subtracting the
wide H$\alpha$ wings excellently approximated by $\Delta\lambda^{-2}$. 
Because H$\alpha$
is too strong and He~II~$\lambda$~6527 is too weak, we present the results 
in two panels. In the upper panel, we show the Gaussian fits to 
He~II~$\lambda$~6527 and H$\alpha$ by dotted lines, 
where all the emission lines are nicely fitted 
with a single Gaussian. The peak values are $f_0=1.7\times 10^3$ for
He~II~$\lambda$~6527 and $f_0=4.8\times 10^4$ for He~II~$\lambda$~6560.
The lower panel shows the detailed result around He~II~$\lambda$~6560.}
\end{figure}

\begin{figure}
\plotone{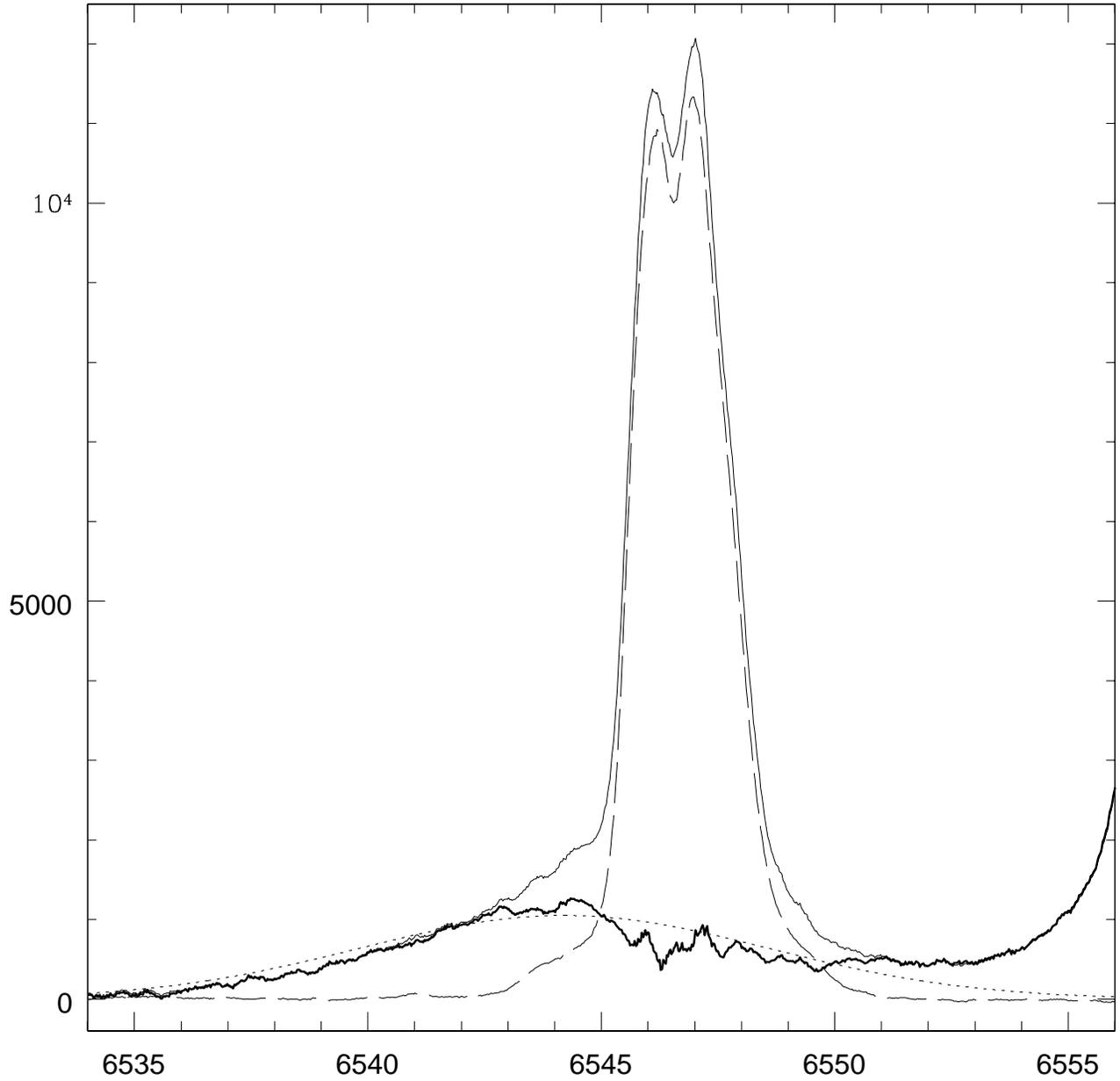} \caption{
The Raman scattered He~II 6545 isolated by deblending from the 
[N~II]~$\lambda$~6548
emission line. Because forbidden lines are optically thin and 
[N~II]~$\lambda$~6548
and [N~II]~$\lambda$~6584 originate from the same excited level, 
the profiles are
exactly same and differs only in strength by a factor of 3. The dashed
line shows the [N~II]~$\lambda$~6584 emission line divided by 3 and translated 
to the location of [N~II]~$\lambda$~6548. The dotted line is a single 
Gaussian fit
to the Raman scattered He~II 6545 feature. The peak value 
$f_0=1.0\times 10^3$ and the width is  6.4 times that
of the He~II~$\lambda$~6527 and He~II~$\lambda$~6560, which is exactly 
expected from the incoherency of Raman scattering.}
\end{figure}
\begin{figure}
\plotone{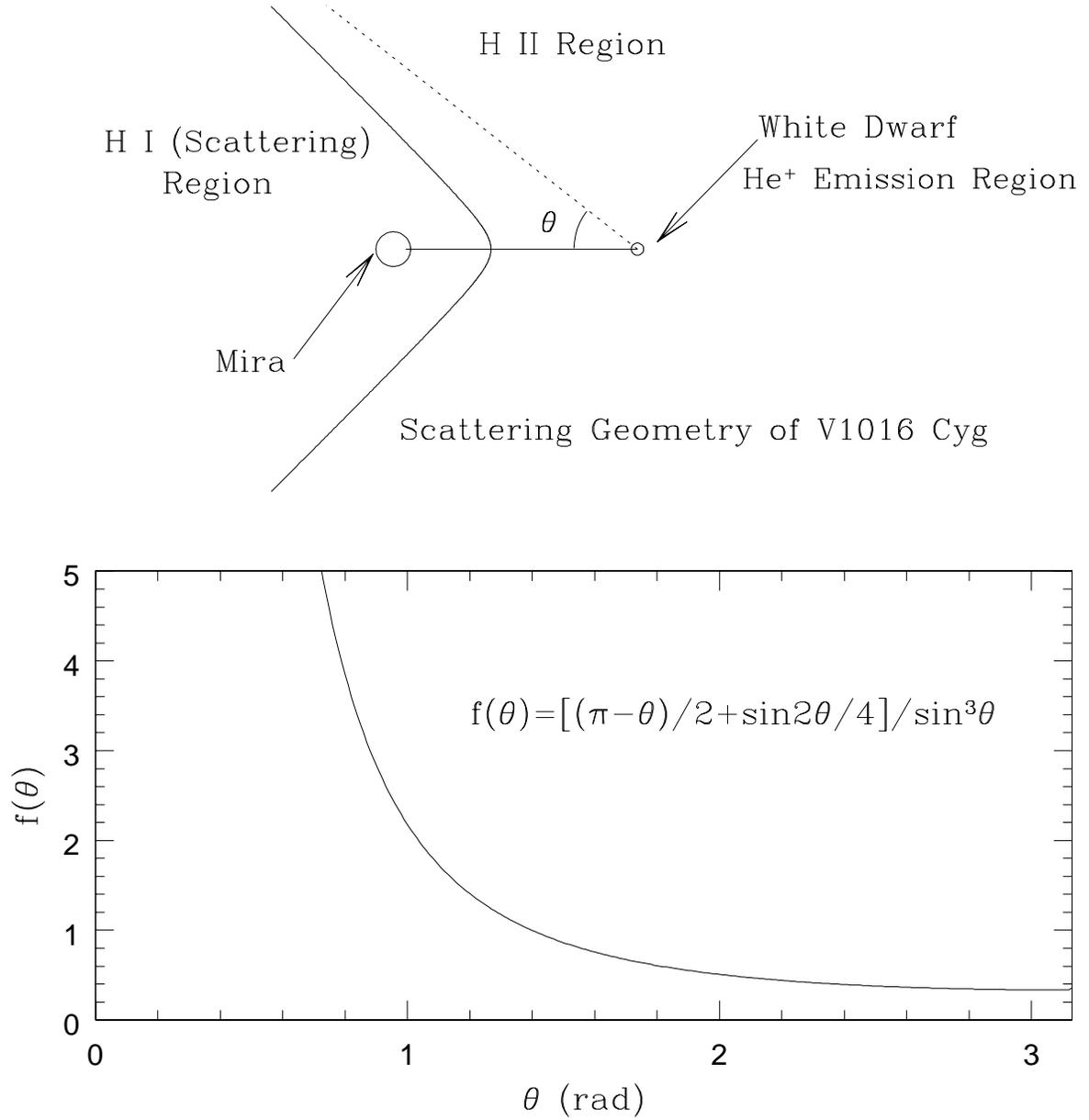} \caption{
The scattering geometry of V1016 Cyg and the ionization of the stellar
wind from the Mira. Due to the strong UV radiation coming from the
mass accreting white dwarf, a significant part of the slow stellar wind
will be ionized. The thick curve shown in the upper panel represents
the ionization front dividing the neutral region around the giant and
ionized region formed in the part where the white dwarf resides.
The low panel shows the behavior of the function $f(\theta)$ defined
in the text.  }
\end{figure}

\end{document}